# Guest Editorial: On-chip mid-infrared and THz frequency combs for spectroscopy

Giacomo Scalari [1], Jérôme Faist [1,], Nathalie Picqué [2]

1. ETH Zurich, Institute of Quantum Electronics, Auguste-Piccard-Hof 1, Zurich 8093, Switzerland
2. Max-Planck Institute of Quantum Optics, Hans-Kopfermannstr. 1, 85748 Garching, Germany

Frequency combs, spectra of phase-coherent equidistant lines often generated by mode-locked femtosecond lasers, have revolutionized time and frequency metrology [1, 2]. In recent years, new frequency comb lasers, of a high compactness or even on-chip, have been demonstrated in the mid-infrared and THz regions of the electromagnetic spectrum. They include electrically pumped quantum cascade and interband cascade semiconductor devices, as well as high-quality factor microresonators. These new comb generators open up novel opportunities for spectroscopy of molecular fingerprints over broad spectral bandwidths: their small form-factor promises chip-scale spectrometers, while their large line spacing simplifies the resolution of the individual comb lines with simple spectrometers and provides short measurement times, with an intriguing potential for time-resolved spectroscopy in the condensed phase. This Letter summarizes the recent advances in this rapidly developing domain of on-chip combs for broadband spectroscopy and discusses some of the challenges and prospects. While some review articles already provide a perspective into some aspects of chip-based frequency comb generators [3-5], of mid-infrared frequency comb synthesizers [6] and of the applications of frequency combs to spectroscopy [7], the growing and rapidly advancing interest in molecular sensing using quantum cascade and microresonator-based on-chip frequency combs motivates this Letter.

An optical frequency comb is an electromagnetic spectrum consisting of equally spaced phase-coherent laser lines, whose equidistance is guaranteed by non-linear optical processes involving typically three or four waves. As a result, the frequency $f_n$ of an individual line of index n can be expressed as $f_n = f_{ceo} + n\, f_{rep}$, where the repetition frequency $f_{rep}$ and the carrier-envelope offset frequency $f_{ceo}$ are radio frequencies. The fluctuation and noise of all the individual lines (Fig.1(a)) are entirely characterized by two parameters, e.g. by the noise of the repetition frequency $f_{rep}$ as well as that of the carrier-envelope offset frequency $f_{ceo}$. The comb possesses a fixed phase relation between the individual spectral modes such that, in the time domain, the output intensity waveform is periodic. However the latter is not bound to be a train of pulses but can have quite an arbitrary waveform. If the two parameters $f_{rep}$ and $f_{ceo}$ are known, then the optical frequencies of all the comb lines are also known. While the repetition frequency $f_{rep}$ is easy to measure with a fast photodiode, the determination of $f_{ceo}$ is more involved. A frequency comb synthesizer is called self-referenced [8, 9] if detection of $f_{ceo}$ is achieved though the measurement of a beat note between two consecutive harmonics (or subharmonics) of comb lines. In such a case, the comb provides a direct link between radio and optical frequencies and enables absolute frequency measurements. Self-referencing is facilitated [10] when the frequency comb spans more than one octave, which enables the implementation of the so-called *f-2f* scheme (Fig.1b). More specifically, the *f-2f* self-referencing technique consist in beating the frequency-doubled low-frequency end of the spectrum (frequencies that can be written $2(n f_{rep} + f_{ceo})$) and the high-frequency end of the spectrum (frequencies that can be expressed as $(2n f_{rep} + f_{ceo})$) The resulting beat note provides $f_{ceo}$.

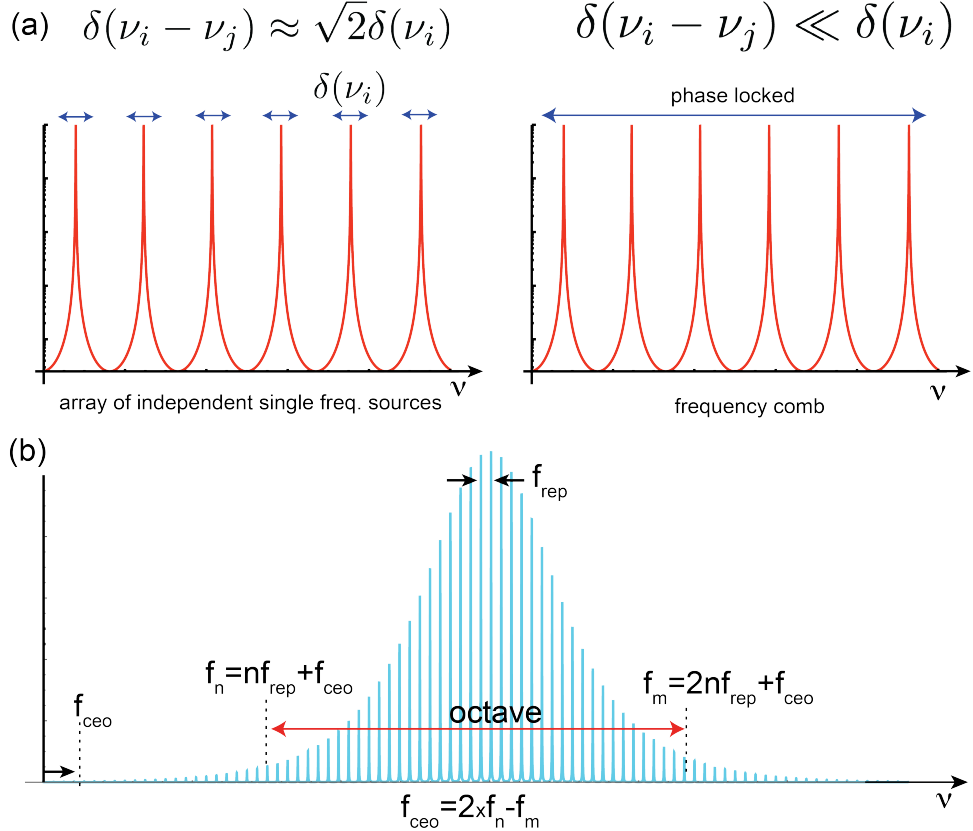

Figure1. (a): the line-to-line noise $\delta(\nu_i-\nu_j)$ of an array of independent sources each one characterized by a noise $\delta(\nu_i)$ is much larger than that of a phase-locked frequency comb. (b): octave-spanning frequency comb: the carrier envelope offset frequency $f_{ceo}$ can be extracted (and thus controlled) by frequency doubling the red part of the spectrum $f_n$ and extracting the beating with the high frequency part $f_m$ as $2xf_n-f_m=2(nf_{rep}+f_{ceo})-nf_{rep}-f_{ceo}=f_{ceo}$, implementing the so-called *f-2f* self-referencing scheme [9, 10].

For more than fifteen years, frequency combs have been harnessed for new purposes -beyond frequency metrology- where their spectral structure of discrete and accurate lines brings some original benefits. As light sources and/or spectrometers, they directly probe an atomic or molecular sample and measure its broad spectrum. Spectroscopy of molecules, in the gas phase and the condensed phase, in the THz (60-250 μm), far-infrared (20-60μm) and mid-infrared (2-20μm) spectral regions, is one of the major applications. Here, most molecules have strong rotational and ro-vibrational transitions, enabling fundamental spectroscopy and sensing. The advent of on-chip frequency combs in the molecular fingerprint region holds much promise for new approaches to the field of frequency-comb spectroscopy, beyond the obvious advantage of compactness. The general architecture for on-chip frequency combs is sketched in Fig.2.

Quantum cascade lasers are semiconductor lasers based on intersubband transitions in quantum wells that can be tailored to emit in the mid-infrared and Terahertz frequency ranges [11, 12]. Recently [13], it was shown that engineering the cavity dispersion in these devices could allow stable optical frequency comb operation [14, 15]. Moreover, the very short upper state lifetime ($\tau \sim 0.6ps$) of the quantum cascade laser active region is both responsible for the resonant, Terahertz-wide bandwidth four-wave mixing that locks the mode equidistance and for enforcing a comb operation in which the output emission is strongly frequency modulated while its output intensity remains relatively constant during a period. Because they are electrically driven, chip-based, and emitting directly in the wavelength range of interest, these devices have attracted a lot of attention recently [13]. Devices operating in the 8-10 [13, 16, 17] and 4-5 µm [18, 19] wavelength range have been demonstrated with 100-$cm^{-1}$ bandwidth and up to one Watt of optical continuous wave power thanks to progress in waveguide and active region designs. The operating range of QCL combs was extended to the THz range too [15, 20-24], where, owing to the intrinsic broadband nature of the waveguides, octave spanning emission has been reported [25] with the relative bandwidth coverage in comb operation as high as $\Delta f/f$= 36% [26]. It has to be noted that operation of THz QCLs is for the moment limited to cryogenic temperatures, even though promising results are coming from on-chip THz DFG with Mid-IR devices [27].
Moreover, experiments have shown that these devices are well adapted for spectroscopy, as their linewidth is Schawlow-Townes limited for short time scales [28], the equidistance between the modes is demonstrated to be better than 1 part per $10^{13}$ [51], and they do not present excess amplitude noise. An important feature of the QCL combs is the possibility to detect and control the repetition frequency of the laser directly from the bias line that powers the device [13]. This feature arises again thanks to the ultrafast electron dynamics of the upper state. In fact, the strong photon driven transport also makes these devices also usable as ultrafast detectors [29]. A challenge is to achieve devices with broader bandwidth comb operation, as systems would clearly benefit from 200-300$cm^{-1}$ bandwidths and possibly lower electrical dissipations than the current 10 W. Since quantum cascade laser do not operate with single pulses, nonlinear optics tools such as spectral broadening using non-linear waveguides is more challenging than with conventional mode-locked lasers
Interband quantum cascade lasers [30, 31] rely on a cascade of interband transitions connected by tunnel junction. These devices have demonstrated recently low-dissipation comb operation in the short wavelength part of the mid-infrared (from 3. 6µm approximately) [32-35] although data are lacking on the stability and coherence. Because they rely on interband transitions with a much longer lifetime, their characteristics are expected [33, 34] to be closer to the ones of near-infrared semiconductor mode-locked lasers than to quantum cascade lasers.
Both the quantum cascade and interband comb lasers are chip-based, and therefore involve short cavities. They naturally operate with a relatively high repetition frequency, on the order of 10 GHz or greater through harmonic mode-locking [36]. Such a high repetition frequency may be difficult to reach using conventional mode-locked lasers systems. One resulting advantage is a high power per line.

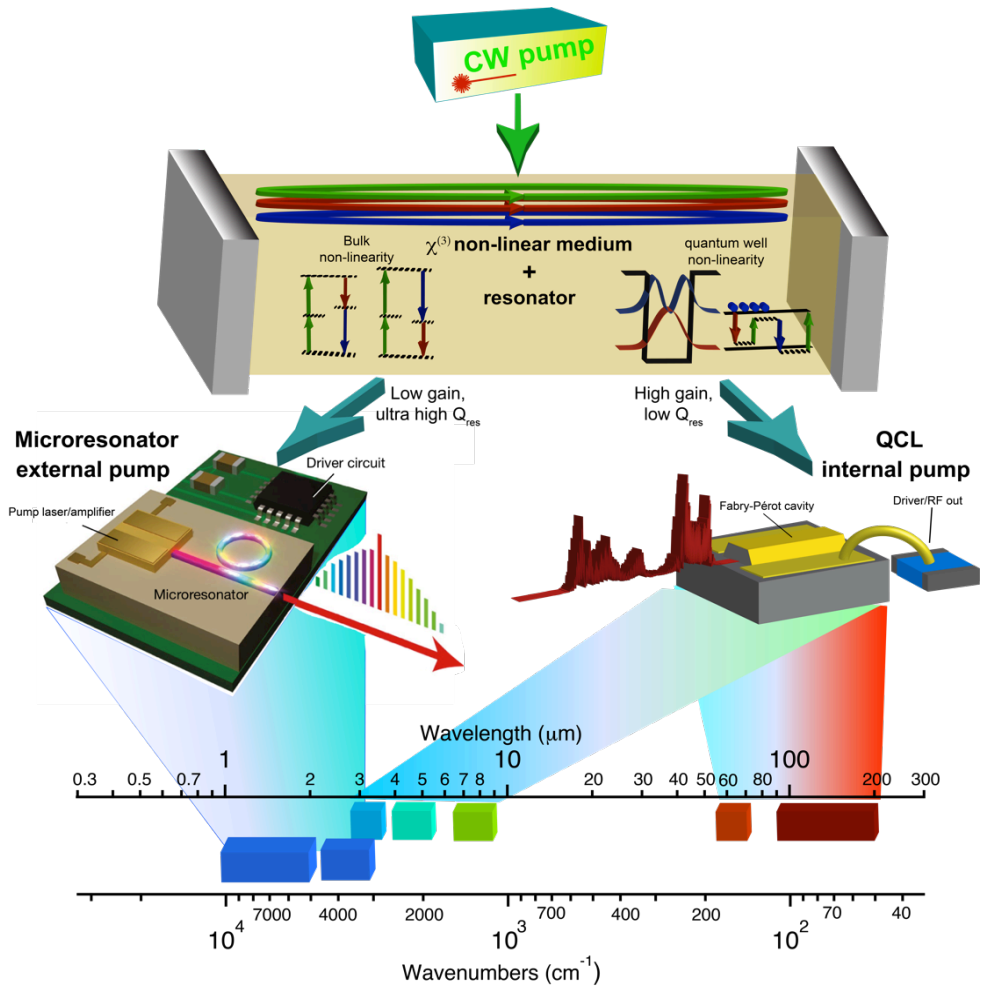

Figure 2. On-chip frequency comb architectures and spectral coverage. (part of the image is a reprint from ref. [37])

Frequency combs can also been generated in high quality factor microresonators pumped by a narrow linewidth, continuous-wave laser [38]. The non-resonant Kerr non-linearity of the microresonator leads to parametric-oscillation-driven cascaded four-wave mixing or even to the formation of dissipative solitons. Chip-based resonators have been demonstrated using a plurality of platforms, most notably $SiO_2$ toroidal resonators [39], CMOS-compatible silicon nitride [40] and silicon [41] waveguides. The technology has become mature in the telecommunication region, where generation of solitons is now achieved with several dispersion-engineered platforms [42]. A battery-operated device has even been reported [37], showing great potential towards fully integrated synthesizers with a very small footprint. The development of microresonator-based frequency combs in the mid-infrared region is challenging and the field is still in its infancy. Nevertheless, some early proofs-of-principle using crystalline [43] and on-chip silicon [41] and silicon nitride [40] microresonators showed frequency combs centered at 2.5 μm. More recently, mode-locked operation leading to broad spectra spanning from 2.4–4.3 μm has been achieved with a silicon platform [44]. The extension towards longer wavelengths faces the difficulties of fabricating low-loss platforms [45], whereas interband or intersubband cascade lasers could potentially provide efficient pump lasers [46]. Broad-spectral bandwidth mode-locked microresonators [5] usually have a repetition frequency higher than 100 GHz, which is inaccessible to solid-state bulk and fiber-doped mode-locked lasers. Such a high repetition frequency may be challenging to detect, especially in the mid-infrared region. Furthermore, a particular feature of microresonator-based combs, which currently restricts the applications to spectroscopy and the possibility of self-referencing by directly beating two harmonics of comb lines, is the very strong line-to-line variation in power. A promising aspect of silicon and silicon nitride mid-infrared platforms is the possibility to engineer the dispersion, enabling e.g. to spectrally broaden within the resonator through dispersive waves. Spectral broadening may also be achieved in dispersion-engineered on-chip nonlinear waveguides. Octave-spanning phase-coherent spectra have been demonstrated with silicon platforms [47-49] pumped by external mid-infrared optical parametric oscillator femtosecond systems and with silicon nitride waveguides pumped by a fiber laser emitting in the telecommunication region [50]. Chalcogenide chip-based waveguides are also powerful for mid-infrared supercontinuum generation [51], though the phase coherence of a broadened comb has not been experimentally verified. Even if mode-locked microresonators generate femtosecond pulses, their high repetition frequency implies a low pulse energy, which may render efficient spectral broadening and other nonlinear processes in a waveguide challenging.

Other emerging approaches to compact or on-chip systems harness external cavity mode-locked semiconductor lasers [52] and $\chi^2$ nonlinear processes in electro-optic modulators [53] and such techniques might be extended to the mid-infrared range in the future.

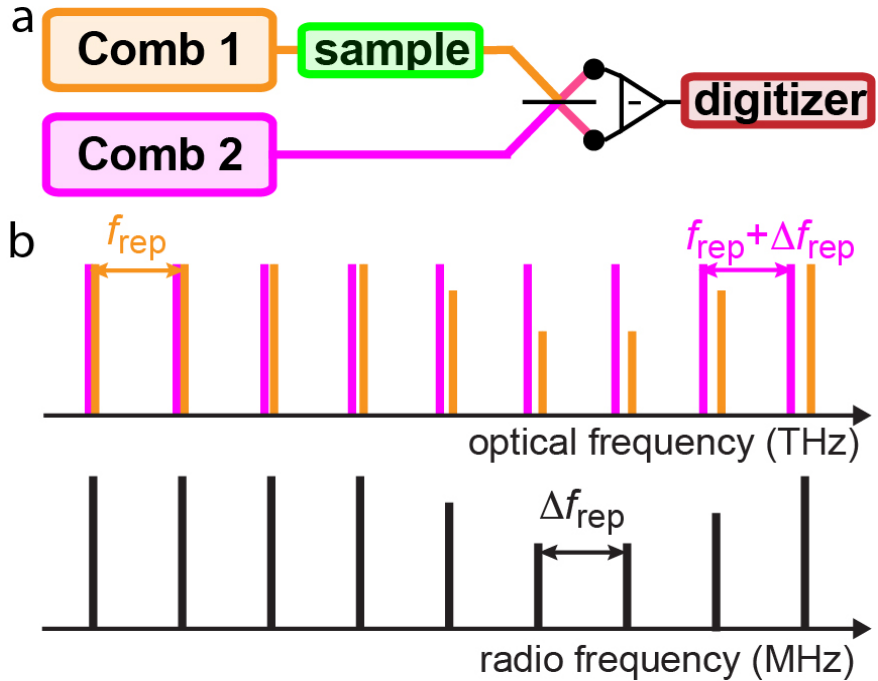

Figure 3. a. Sketch of a set-up for dual-comb linear absorption spectroscopy: a frequency comb interrogates an absorbing sample and beats with a second comb, of slightly different line spacing, on a fast photodetector. The interference signal is measured as a function of time. b. Frequency domain representation of the measurement principle. Beat notes between pairs of optical comb lines, one from each comb, generate a radio-frequency comb that can be measured with digital electronics.

The spectroscopy instrument that has so far aroused the highest interest is the dual-comb spectrometer (Fig.3a). Dual-comb spectroscopy is a technique of Fourier-transform spectroscopy, where the time-domain interference (called an interferogram) between two combs of slightly different line spacing is monitored and Fourier-transformed to reveal the spectrum. The frequency-domain picture of its physical principle is shown in Fig.3b. Owing to interference between two frequency combs of slightly differing line spacing $f_{rep}$ and $f_{rep}+\Delta f_{rep}$, pairs of optical lines, one from each comb, produce radio-frequency beat notes on the detector. The frequencies of the optical spectrum are down-converted to $\Delta f_{ceo}+ n\Delta f_{rep}$, where $\Delta f_{ceo}$ is the difference of carrier-envelope offset frequencies of the two combs and n an integer. The radio-frequency comb of line spacing $\Delta f_{rep}$, formed by the beat signals is easily measurable by standard electronics tools. Details may be found in the review article [7]. Besides spectroscopy, the dual-comb technique proves also useful for characterizing frequency comb generators and for other applications of interferometry. In the THz and mid-infrared regions (for early references, see e.g. [54-56]), dual-comb spectrometers are particularly attractive because they measure -without moving parts- spectra as broad as the span of the combs, on a single photo-detector. Multiplex recording grants an exceptional consistency to the spectral measurements and circumvents expensive detector arrays. Resolving the comb lines sets the spectral resolution to the line spacing $f_{rep}$. The resolution may be further improved, by interleaving spectra where the positions of the comb lines are sequentially stepped. Dual-comb spectroscopy is however prone to numerous artifacts that can strongly degrade the spectra. For instance, during the time of the measurement, the two combs must be kept coherent. With fiber-doped mode-locked laser systems, this has been experimentally achieved by locking the two combs to common continuous-wave lasers of very narrow line-widths [57] or by feed-forward control of the relative carrier-envelope offset [58, 59]. Long coherence times have not yet been experimentally demonstrated with on-chip combs, as the technology is still emerging. Alternatively, relative fluctuations of the combs can be corrected by analog [60] or digital [61] treatment. *A-posteriori* correction algorithms have been successfully developed with quantum and interband cascade lasers [32, 62]. Assuming that artifacts (due to e.g. combs not mutually coherent or detector nonlinearities) are minimized, the accuracy in the spectra depends on the stabilization of the combs. A frequency comb can be stabilized to a radio-frequency clock, through the self-referencing technique -not yet accessible to THz and mid-infrared on-chip combs- or to stable continuous-wave lasers. So far, on-chip dual-comb systems have been left free running and do not achieve precision measurements.

The initial publications with semiconductor combs have shown convincing potential for gas spectroscopy at 8.5 μm [63], 7.9 μm [64] 7.1 μm [65] and 3.2 μm [66] with a span up to one THz (35 cm$^{-1}$), for multi-heterodyne beats at 10 μm [67] and at 5 μm [19], and for control and tuning of the line positions at 7.5 μm [68] using micro-heaters. With wavelength modulation techniques [64] or electro-optic sampling [69], the sensitivity to weak absorptions is improved. The detection of the multi-heterodyne signal in the bias current of one of the THz combs provides intriguing prospects for integrated compact spectrometers [29]. However, for gas-phase spectroscopy, the line spacing of semiconductor combs is too large compared to the width of the molecular lines. Spectral interleaving, which may decrease the precision if the sample varies from one recording to the next, is still challenging, as it requires accurate and fast control of the comb line positions. Gas-phase dual-comb spectroscopy with semiconductor combs in the mid-infrared and THz ranges is thus still largely unexplored and open to new insights and improvements. The high power per line of on-chip combs could for instance benefit nonlinear spectroscopy. A key challenge remains the increase of the spectral span.

A unique potential of on-chip comb sources, which can hardly be tackled with conventional mode-locked laser systems, addresses rapid spectroscopy of condensed matter, where the width of the bands is generally broader than 100 GHz. In a dual-comb system, an interferogram of a resolution equal to the line spacing $f_{rep}$ recurs every $1/\Delta f_{rep} >= 2\ \Delta f/f_{rep}^2$, where $\Delta f$ is the span of the laser spectrum. For example, for a span $\Delta f$=75 THz, the recording of a single-shot interferogram leading to resolved comb lines of $f_{rep}$=100 MHz requires at least 15 ms, whereas with combs of $f_{rep}$=100 GHz, it shortens to 15 ns! This exhilarating perspective for time-resolved spectroscopy of single events in the molecular fingerprint region could open up new opportunities in chemistry and biology, where Michelson-based Fourier transform spectroscopy has been routinely used for decades. The large spectral intensity variations -from one line to the next- in on-chip combs, the poor dynamic range of fast digitizers and the need for broad spectral spans make the prospect dauntingly challenging though. Nevertheless, the first proof-of-concept demonstration of mid-infrared microresonator-based dual-comb spectroscopy shows [70] vibrational spectra of liquid acetone spanning from 3.10 to 2.90 μm at 127-GHz (4.2 cm$^{-1}$) resolution, measured within 2 µs. Recently, time-resolved dual-comb spectroscopy using quantum cascade lasers between 8.44 and 8.05 µm has investigated [71] the photocycle of bacteriorhodopsin, with a µs time resolution and a spectral resolution of 130 GHz (4.3 cm$^{-1}$). Repeating the reaction only tens to hundreds of times proved sufficient to extract relevant kinetics. Another application that may take advantage of short measurement times is spectral mapping, where a spatially inhomogeneous sample is raster-scanned and a dual-comb spectrum is measured at each position [72]. Very recently a THz quantum cascade dual-comb system, centered at 88 µm, has investigated, with a spatial resolution of 0.5 mm and 81x53 pixels, a solid containing three different absorbers [73].

In another approach to gas-phase spectroscopy [74], the light of a broad spectral-bandwidth mid-infrared mode-locked microresonator comb of a large line spacing (127 GHz, 4.2 cm$^{-1}$) is analyzed by a commercial Fourier transform spectrometer. The position of the comb lines is tuned per step of 80 MHz (2.4 10$^{-3}$ cm$^{-1}$), enhancing the sampling interval by spectral interleaving. This points to yet-unexplored possibilities. In the future, the comb lines could be resolved by on-chip low-resolution spectrometers that may rely on Fourier transform interferometers based on micro-mechanical moving mirrors or on stationary waves [75], or on gratings that would not even require crossed dispersion.

On-chip frequency comb synthesizers are under active development in the THz and mid-infrared regions. Small footprint, electrical pumping, large line spacing, high power per line and high brightness are characteristics already afforded by some of these generators. Significant challenges aiming at achieving broad spectral spans, accurate frequency control and flat spectral intensity distribution motivate ongoing technical developments. Through first proofs of concept in gas-phase and condensed-phase spectroscopy, quantum (interband) cascade and microresonator frequency comb generators show an intriguing potential for molecular spectroscopy and sensing. When harnessed in dual-comb spectroscopy, their large line spacing might offer new keys to analytical chemistry for the study of the kinetics of chemical reactions in the liquid phase, while their high power per comb line might efficiently generate nonlinear phenomena at the sample, useful for e.g. precision spectroscopy of molecules. Such exciting prospects attract a growing number of scientists and may stimulate their creativity for overcoming the instrumental challenges associated with the optimization of dual-comb interferometers. The exploration of other schemes of spectrometry is likely to lead to powerful instruments too. Such efforts might even culminate in chip-scale spectrometers in the molecular fingerprint region, able to tackle modern defies in physics, environmental sciences, chemistry or biomedicine.

G.S. and J.F. would like to gratefully acknowledge financial support from ERC consolidator CHiC grant 724344 and from SNF through grant 200020_178942.